\documentstyle[twoside,fleqn,espcrc2,epsf]{article}
\newcommand{\be}{\begin{equation}}
\newcommand{\ee}{\end{equation}}
\newcommand{\bea}{\begin{eqnarray}}
\newcommand{\eea}{\end{eqnarray}}

\def\bfnabla{\mbox{\boldmath $\nabla$}}
\def\bfSigma{\mbox{\boldmath $\Sigma$}}
\def\bfsigma{\mbox{\boldmath $\sigma$}}
\def\bfPi{\mbox{\boldmath $\Pi$}}
\def\lQ{\Lambda_{\rm QCD}}
\def\al{\alpha}
\def\als{\alpha_{\rm s}}
\def\siml{{\ \lower-1.2pt\vbox{\hbox{\rlap{$<$}\lower6pt\vbox{\hbox{$\sim$}}}}\ }} 

\def\lla{\langle\!\langle}
\def\rra{\rangle\!\rangle}
\newcommand{\Appendix}[1]%
    {%
     \section{#1}%
      }

\begin{document}
\def\siml{{\ \lower-1.2pt\vbox{\hbox{\rlap{$<$}\lower6pt\vbox{\hbox{$\sim$}}}}\ }} 
\def\bfnabla{\mbox{\boldmath $\nabla$}}
\def\bfSigma{\mbox{\boldmath $\Sigma$}}
\def\bfsigma{\mbox{\boldmath $\sigma$}}
\def\als{\alpha_{\rm s}}
\def\al{\alpha}
\def\lQ{\Lambda_{\rm QCD}}
\def\vs{V^{(0)}_s}
\def\vo{V^{(0)}_o}
\def\bfgamma{\mbox{\boldmath $\gamma$}}
\def\bfnabla{\mbox{\boldmath $\nabla$}}
\def\bfSigma{\mbox{\boldmath $\Sigma$}}
\def\bfsigma{\mbox{\boldmath $\sigma$}}
\def\bfxi{\mbox{\boldmath $\xi$}}
\def\bfepsilon{\mbox{\boldmath $\epsilon$}}
\def\bfPi{\mbox{\boldmath $\Pi$}}
\def\lQ{\Lambda_{\rm QCD}}
\newcommand{\one}{1\!\!{\rm l}}
\newcommand{\onec}{1\!\!{\rm l}_c}
\newcommand{\ones}{1\!\!{\rm l}_s}

\def\al{\alpha}
\def\als{\alpha_{\rm s}}
\def\siml{{\ \lower-1.2pt\vbox{\hbox{\rlap{$<$}\lower6pt\vbox{\hbox{$\sim$}}}}\ }} 
\def\simg{{\ \lower-1.2pt\vbox{\hbox{\rlap{$>$}\lower6pt\vbox{\hbox{$\sim$}}}}\ }} 
\newcommand{\MS}{\overline{\rm MS}}
\def\lla{\langle\!\langle}
\def\rra{\rangle\!\rangle}
\newcommand{\ttbs}{\char'134}
\newcommand{\AmS}{{\protect\the\textfont2 A\kern-.1667em\lower.5ex\hbox{M}\kern-.125emS}}

\title{Inclusive Quarkonium Decays}

\author{Nora Brambilla\address{Dipartimento di Fisica, U. Milano I and INFN Milano, \\ 
Via Celoria 16, 20133 Milano, Italy}}

\begin{abstract}
I review recent progress in the calculation of inclusive quarkonium decays
in the framework of QCD nonrelativistic effective field theories and in relation 
to the  experimental measurements.
\end{abstract}

\maketitle

\section{INTRODUCTION}
Inclusive quarkonium decays, more precisely
inclusive annihilation decay rates 
of heavy quarkonium states into light hadrons
(hadronic decays)  and photons and lepton pairs (electromagnetic decays), 
were among the earliest calculations of perturbative QCD.
There, it was assumed that the decay rate of the quarkonium state 
factored into a short distance part, calculated as the annihilation rate 
of the heavy quark and antiquark and given in terms of  $\als(m)$, 
and a long distance nonperturbative part given in terms of 
 the quarkonium wave function  (or its derivatives) evaluated at the origin.
Explicit calculations at next to leading order in $\als$ 
in perturbation theory for $S$- and $P$-wave decays supported  
 the factorization assumption which could not however be proved on general grounds for higher orders of perturbation theory. Indeed, in the case 
of $P$-wave decays into light hadrons, it turned out that at order $\als^3$ 
 the factorization was spoiled by logarithmic infrared divergences 
\cite{pqcd}. 
The same problem appeared in relativistic corrections to the annihilation 
decays of $S$-wave states \cite{pqcd}.

This problem has been solved \cite{nrqcd1,nrqcd2}   and new predictions have been 
obtained \cite{Mont,long}  with the introduction of non-relativistic effective field 
theories (EFTs) of QCD,  that 
has put our description of these systems
on the solid ground of QCD. 

The reason for which the EFT approach is so successfull for heavy quarkonium 
is the following. Heavy quarkonium, being a non-relativistic bound state, 
is characterized by a hierarchy of energy scales $m$, $mv$ and $mv^2$, where $m$ is the heavy-quark 
mass and $v\ll 1$ the relative heavy-quark velocity.
A hierarchy of EFTs may be constructed by systematically integrating out 
modes associated to these energy scales in a matching procedure that 
enforces the complete equivalence between QCD and the EFT at a given 
order of the expansion in $v$ and $\als$.

 Integrating out degrees of freedom 
of energy $m$, which for heavy quarks can be done perturbatively, leads to  
non-relativistic QCD (NRQCD)\cite{nrqcd1,nrqcd2}. This EFT still contains the lower 
energy scales as dynamical degrees of freedom. In the last years, the problem 
of integrating out the remaining dynamical scales of NRQCD has been addressed 
by several groups and has reached  a good level of 
understanding (a  list of references can be found in \cite{reveft}). 
The EFT obtained by subsequent matchings from QCD, where only the
lightest degrees of freedom of energy $mv^2$ are left dynamical, 
is called  potential NRQCD, pNRQCD \cite{Mont,long}. This EFT is close to a
quantum-mechanical description of the bound system and, therefore, as simple. 
It has been systematically explored in the dynamical
regime $\lQ \siml mv^2$ in \cite{long,logs,logss} and in the regime $mv^2 \ll \lQ \siml mv$ 
in \cite{long,M12,sw}. The quantity $\lQ$ stands for the generic scale of 
non-perturbative physics.

The EFT approach  made it possible, in the case of several
observables, among which the inclusive decay widths, to achieve a rigorous factorization between 
 the high-energy dynamics encoded into matching coefficients 
calculable in perturbation theory and the non-perturbative QCD dynamics 
encoded  into few well-defined matrix elements.  
Systematic
improvements are possible, either by calculating higher-order corrections 
in the coupling constant or by adding higher-order operators.
In this way the prediction of inclusive decays are put in direct 
relation to QCD and the theoretical uncertainties may be consistently estimated.

From the experimental side new data have recently been produced 
for heavy-quarkonium observables. Measurements relevant to the
determination of heavy-quarkonium inclusive decay widths have come from 
Fermilab (E835)\cite{E835}, BES\cite{BES}, CLEO\cite{CLEO1,CLEO2} and Belle\cite{Belle} 
\footnote{For updates see also \cite{qwg}.}
and demand  accurate QCD theoretical predictions. 
On the other hand the inclusive decays of heavy quarkonium may provide 
competitive information on $\als$ at the scale $m$ once the 
theoretical and experimental errors are under control.

In the following I will review recent progress in our theoretical understanding
of inclusive and electromagnetic heavy-quarkonium decays. I will recall the
NRQCD factorization results in Sec. \ref{secnrqcd} and the further
simplification achieved within the  
pNRQCD  factorization 
 in Sec. \ref{secpnrqcd}. The presented pNRQCD formulas apply 
to quarkonia that fulfil $mv \simg \lQ \gg mv^2$. 

\vskip -0.7truecm

\section{NRQCD}
\label{secnrqcd}
NRQCD is the EFT obtained by integrating out 
the hard scale $m$. This $m$ being larger than the scale 
of non-perturbative physics, $\lQ$, the matching to NRQCD can be done 
order by order in $\als$. Hence, the NRQCD Lagrangian can be written 
as a sum of terms like $ f_n \, O^{(d_n)}_n/m^{d_n-4}$, ordered in powers of $\als$ 
and $v$. More specifically, the Wilson coefficients 
$f_n$ are series in $\als(m)$ and encode the ultraviolet physics that  
has been integrated out from QCD.
 The operators $O^{(d_n)}_n$ of dimension $d_n$
describe the low-energy dynamics and are counted in powers of $v$.
Heavy quarkonium inclusive decays are controlled by the imaginary part 
of the NRQCD Hamiltonian, i.e. the 
imaginary part of the Wilson coefficients of the 4-fermion operators 
($O^{(d_n)}_n = \psi^\dagger K_n  \chi \chi^\dagger K^\prime_n \psi$) 
in the NRQCD Lagrangian. The NRQCD factorization formula for 
quarkonium (H) inclusive decay widths into light hadrons (LH) reads \cite{nrqcd2}
\begin{eqnarray}
\!\!\!\!\!\!\!\!\!\!\!\!\!\!\!\!\!& &\Gamma({\rm H}\to{\rm LH}) = 
\sum_n {2 \, {\rm Im} \, f_n \over m^{d_n - 4}}
\; \langle {\rm H} | \psi^\dagger K_n  \chi \chi^\dagger K^\prime_n \psi |{\rm H} \rangle.
\label{fac1}
\end{eqnarray}
The 4-fermion operators are classified with respect to their rotational 
and spin symmetry (e.g. $O(^{2S+1}S_J)$, $O(^{2S+1}P_J)$, ...) and of their 
colour content (octet, $O_8$, and singlet, $O_1$, operators).
Singlet operator expectation values may be easily related to the square 
of the quarkonium wave functions (or derivatives of it) at the origin. 
These are unknown non-perturbative parameters.


Let us make a concrete example by considering the P-wave inclusive decays.

In NRQCD the $P$-wave inclusive decay width for the $S=0$ ($h$) and 
$S=1$ ($\chi$) quarkonium states is given at leading (non-vanishing) 
order in $v$ (which is $mv^5$) by \cite{nrqcd2}:
\begin{eqnarray}
& & 
\!\!\!\!\!\!\!\!\!\! 
\Gamma(h  \to {\rm LH})
= {9 \; {\rm Im \,}  f_1(^1P_1)  \over \pi m^4}   
\Big| {R_P'} \Big|^2
\nonumber  \\
& & 
+ {2  {\rm Im}  f_8(^1S_0)  \over m^2} 
\langle h | O_8 (^1S_0) | h \rangle, 
\label{Pnrqcd1}
\\  
& &
\!\!\!\!\!\!\!\!\!\!
\Gamma(\chi_{J}  \to {\rm LH})
 = 
{9  {\rm Im}  f_1(^3P_J)  \over \pi m^4}   
\Big| {R_P'} \Big|^2 \nonumber \\
& &
+ {2  {\rm Im }  f_8(^3S_1)  \over m^2}  
\langle h | O_8(^1S_0 ) | h \rangle,  {\rm for}  J=0,1,2~~~~
\label{Pnrqcd2}
\end{eqnarray}
where $R_P'$ is the derivative of the radial $P$-wave function at the origin.
We stress that, according to the power counting of NRQCD, the 
octet  contribution $\langle h | O_8 (^1S_0) | h \rangle$ 
is as relevant as the singlet contribution \cite{nrqcd2}.
This octet contribution reabsorbs the dependence 
on the infrared cut-off $\mu$ of the Wilson coefficients ${\rm Im \,} f_1(P)$
solving the problem mentioned in the introduction.
From the above equations, we see that in NRQCD the 8 $P$-wave bottomonium states ($1P$, $2P$) 
and the 4 $P$-wave charmonium states ($1P$), which lie under threshold, depend  
at leading order in the velocity expansion on 6 non-perturbative parameters 
(3 wave functions $+$ 3 octet matrix elements).
The inclusive decays of the heavy quarkonium (either hadronic or
electromagnetic) are usually considered up to, and including, NRQCD
matrix elements of 4-fermion operators of dimension 8. This means
to consider the ${\cal O}(1/m^2,1/m^4)$ local 4-fermion operators of the NRQCD
Lagrangian and the decay rate up to order $mv^5$ 
in the $v$ counting.
 If we
consider that in the bottomonium system in principle 14 $S$- and
$P$-wave states lie below threshold ($\Upsilon(nS)$ and $\eta_b(nS)$
with $n=1,2,3$; $h_b(nP)$ and $\chi_{bJ}(nP)$ with $n=1,2$ and
$J=0,1,2$) and that in the charmonium system this is the case for 8
states ($\psi(nS)$ and $\eta_c(nS)$ with $n=1,2$; $h_c(1P)$ and
$\chi_{cJ}(1P)$ with $J=0,1,2$), all the bottomonium and charmonium
$S$- and $P$-wave decays into light hadrons and into photons or
$e^+e^-$ are then  described in NRQCD up to $v^5$ by 46 unknown
NRQCD matrix elements (40 for the $S$-wave decays and $6$ for the $P$-wave
decays), where we have already used spin symmetry and vacuum saturation
\cite{nrqcd2,sw}.
 These matrix elements have to be fixed either by lattice
simulations \cite{latbo} or by fitting the data \cite{Maltoni}. Only
in the specific case of matrix elements of singlet operators does NRQCD
allow an interpretation in terms of quarkonium wave functions and one
can resort to potential models.

It has been discussed, in particular in \cite{mawang} (but see
also \cite{bope}), that higher-order operators, not considered in the above 
formulas, can be numerically quite relevant. This may be the case particularly 
for charmonium, where $v_c^2 \sim 0.3$, so that relativistic corrections 
are large, and for $P$-wave decays where the above formulas provide,
indeed, only the leading-order contribution in the velocity expansion.
Besides this, precise theoretical predictions are also hampered by uncertainties 
in the NRQCD matrix elements and large corrections in NLO in $\als$.
The convergence of the perturbative series of the four-fermion matching coefficients 
is indeed often bad (for examples see \cite{vairo1}). 
A solution may be provided by the resummation of the 
large contributions in the perturbative series 
coming from bubble-chain diagrams. This analysis has been successfully carried
out in some specific cases in \cite{chen}
but a  general treatment is still missing, in particular in the case of $P$-wave decays.

Here we have counted NRQCD matrix elements by their dimensionality only. 
In fact the power counting of
the NRQCD matrix elements is an open issue. 
In the standard NRQCD power counting,
the octet matrix elements are ${\cal O}(v^4)$ suppressed for $S$-wave decays if
compared with the leading order. This is not so within pNRQCD, assuming the
counting $\lQ \sim mv$, they would only be ${\cal O}(v^2)$-suppressed \cite{sw}. This is
potentially relevant for  $\Gamma(V \rightarrow LH)$ since ${\rm Im}\,
f_1( ^3S_1)$ is ${\cal O}(\als(m))$-suppressed with respect to ${\rm Im} \,f_8( S)$ \cite{sw}. 
In other words, the octet matrix element effects could
potentially be much more important than usually thought for these decays.

In the next section we will show that in the framework of pNRQCD
 it is possible to achieve a noticeable  reduction in the number of the
nonperturbative  parameters and thus 
to formulate  new predictions with
respect to NRQCD.

\vskip -0.7truecm

\section{pNRQCD}
\label{secpnrqcd}

Pushing further the EFT programme for non-relativistic bound states, 
further simplifications occur if we integrate out also soft degrees of freedom.
pNRQCD is the resulting EFT. We will consider pNRQCD under 
the condition $\lQ \gg mv^2$. Then, two situations are possible.
First, the situation when $m v \gg \lQ \gg mv^2$.
In this case the soft scale $m v$ can be integrated out perturbatively. This leads to
an intermediate EFT that contains singlet and octet quarkonium fields 
and ultrasoft gluons as dynamical degrees of freedom.  
The octet quarkonium field and the ultrasoft gluons are eventually 
integrated out by the (nonperturbative) matching to pNRQCD \cite{long}. 
Second, the situation when $\lQ \sim mv$. In this case the (nonperturbative) 
matching to pNRQCD has to be done in one single step \cite{M12}.
Under the circumstances that other degrees of freedom 
develop a mass gap of order $\lQ$ 
the quarkonium singlet field $\rm S$ remains as the only dynamical 
degree of freedom in the pNRQCD Lagrangian, which reads \cite{long,M12,sw}
${\mathcal{L}}_{\rm pNRQCD}= 
{\rm Tr} \,\Big\{ {\rm S}^\dagger \left( i\partial_0 - {\cal H}  \right) {\rm S} \Big
\}$, ${\cal H}$ being the pNRQCD Hamiltonian, to be determined by matching pNRQCD to NRQCD.
The inclusive quarkonium decay width into light hadrons is given by 
\be
  \Gamma ({\rm H}\to{\rm LH}) = - 2\, {\rm Im} \, \langle n,L,S,J| {\cal H}  |n,L,S,J \rangle,
\label{imag}
\ee
where $|n,L,S,J \rangle$ is an eigenstate of ${\cal H}$ with the quantum numbers 
of the quarkonium state $\rm H$.

\subsection{Matching in a $1/m$ expansion}
We consider first the case in which the matching between 
NRQCD and pNRQCD is made within a $1/m$ expansion.
In this case from the matching we obtain schematically:
\begin{eqnarray}
& & \!\!\!\!\!\!\!\!\!\!\!\!\!\!\! {\rm Im} \, {\cal H} = \delta^3({\bf r}) 
\sum_n {{\rm Im} \, f_n \over m^{d_n - 4}} {\cal A}_n 
+ \{ \delta^3({\bf r}), \Delta \}  
\label{imh} \\
& & 
\!\!\!\!\!\!\!\! \times \sum_n {{\rm Im} \, f_n \over m^{d_n - 4}} {\cal B}_n 
+ {\bfnabla}^i\delta^3({\bf r}){\bfnabla}^j
\sum_n {{\rm Im} \, f_n \over m^{d_n - 4}} {\cal C}_n^{ij}
+ \dots,
\nonumber
\end{eqnarray}
where the imaginary part of $f_n$ are  inherited from 
the 4-heavy-fermion NRQCD matching coefficients, 
and ${\cal A}_n$, ${\cal B}_n$, ... are nonperturbative operators, which are 
universal in the sense that they do not depend either on the heavy-quark flavour
or on the specific quantum numbers of the considered heavy-quarkonium state.
Inserting Eq. (\ref{imh}) into (\ref{imag}) and comparing with
Eq. (\ref{fac1}), we see \cite{sw}
 that all NRQCD matrix elements, including the octet
ones, can be expressed through pNRQCD as products of  universal 
nonperturbative factors by the squares of the quarkonium wave functions
(or derivatives of it) at the origin. 
In \cite{sw}  the inclusive decay widths into light hadrons, 
photons and lepton pairs of all $S$-wave and $P$-wave states (under threshold) 
have been calculated up to ${\mathcal O}(mv^3\times (\Lambda_{\rm QCD}^2/m^2,E/m))$ 
and ${\mathcal O}(mv^5)$. A  large reduction in
the number of unknown nonperturbative parameters is achieved and, therefore, 
new model-independent QCD predictions may be obtained. 
The universal nonperturbative parameters are all expressed only in 
terms of gluonic field-strength 
correlators, which may be fixed by experimental data or by lattice simulations.
Thus at the same level of accuracy 
discussed before in NRQCD, $S$- and $P$-wave bottomonium and
charmonium decays are described in pNRQCD, under the dynamical
assumption $\lQ \gg mv^2$ and within a $1/m$ expansion matching, by only 19 nonperturbative
parameters. These are the 13 wave functions
and 6 universal nonperturbative parameters.

This same approach may be useful also for quarkonium production.

\subsection{Contributions from the scale $\sqrt{m\lQ}$ }

Once the methodology to compute the potentials  (real and imaginary contributions) and 
from these the inclusive decays, within a 
$1/m$ expansion in the matching has been developed, the next question that 
appears naturally is  to which extent one can compute the {\it full} potential 
within a $1/m$ expansion in the case $\lQ \gg mv^2$.
It has been recently shown \cite{sqrt} that  new non-analytic terms 
arise due to the three-momentum scale $\sqrt{m\lQ}$. These terms 
can be incorporated into local potentials ($\delta^3 ({\bf r})$ and derivatives of it) 
and scale as half-integer powers of $1/m$. Moreover, 
it is possible to factorize these effects in a model independent 
way and compute them within a systematic expansion in some small parameters.  

These terms are due to the existence of degrees of
freedom, namely the quark-antiquark pair, with relative 
three-momentum of order $\sqrt{m\lQ}$. 
The on-shell energy of these degrees of freedom is of $O(\lQ)$, i.e. 
the same energy scale that is integrated out when computing the standard 
$1/m$ potentials, which corresponds to integrating out (off-shell)
quark-antiquark pairs of three momentum of order $\lQ$. 
Therefore  both degrees of freedom should be 
integrated out at the same time.
Note that the scale$\sqrt{m\lQ}$ fulfils  $\sqrt{m\lQ}$ $\gg \lQ$, 
from which it follows that at this scale we always are in the
perturbative  regime. The matching may be performed in two different ways depending on the 
two situations $mv \gg \lQ$ or $\lQ \siml  mv$ \cite{sqrt}.

The result for the inclusive decays is the following. 
In general, the size of the contributions coming from the scale $\sqrt{m\lQ}$ 
depends on the size of $\als(\sqrt{m\lQ})$ \cite{sqrt}. 
For $P$ decays the leading effect turns out to be $O(m\als/\sqrt{m\lQ})$ 
suppressed with respect to the leading contribution of order $mv^5$
 (we assume $m \als \ll \sqrt{m \lQ}$). 
For the $S$-wave decay widths the leading effect is 
$O\left((\lQ/m) \, (m\als / \sqrt{m\lQ})\right)$ 
suppressed with respect to the leading contribution of order $mv^3$.
All the results fulfil the same factorization properties as
those obtained in \cite{sw} and mentioned in Section 3.1, and thus the nonperturbative parameters
are still encoded into few, only glue dependent, operators.

\vskip -0.7truecm

\subsection{Applications to $P$ decays}
Here we show  how the reduction 
obtained by pNRQCD in the numbers of unknown 
nonperturbative factors makes new 
theoretical predictions possible in the case of $P$-wave inclusive 
quarkonium decays into light hadrons \cite{sw,vairo}.

In pNRQCD the $P$-wave inclusive decay widths are given at leading 
order in $v$  ( which is $v^5$) by \cite{sw}:
\begin{eqnarray}
& &
\!\!\!\!\!\!\!\!\!\!\!\!\!\!\!\!\! \Gamma( h  \to {\rm LH})
= {\Big| { {R_P'}} \Big|^2  \over \pi m^4} 
\left[ 9 \; {\rm Im \,}  f_1(^1P_1) 
+ {{\rm Im }  f_8(^1S_0)  \over 9} 
{\mathcal E} \right]\!,\;\;\;\;
\label{Ppnrqcd1}
\\  
& & \!\!\!\!\!\!\!\!\!\!\!\!\!\!\!\!\!
\Gamma( \chi_{J}  \to {\rm LH})
 = 
{\Big| { {R_P'}} \Big|^2  \over \pi m^4}  
\left[ 9  {\rm Im }  f_1(^3P_J) 
+ {{\rm Im }  f_8(^3S_1)  \over 9} 
{\mathcal E} \right]\!. \;\;\;\;
\label{Ppnrqcd2}
\end{eqnarray}
$ {\mathcal{E}} = \displaystyle {1\over 2} \int_{0}^{\infty} d t \; t^3  
\left\langle g {\bf E}^a(t,{\bf 0}) 
\Phi_{ab}(t,0;{\bf 0}) g{\bf E}^b(0,{\bf 0}) \right\rangle$ is the universal 
nonperturbative (only glue dependent)
 parameter that describes $P$-wave quarkonium decays in pNRQCD. 

By comparing Eqs. (\ref{Pnrqcd1}) and (\ref{Pnrqcd2}) with 
Eqs. (\ref{Ppnrqcd1}) and (\ref{Ppnrqcd2}) we get at leading order in $v$ 
the relation between the octet matrix element of NRQCD and ${\mathcal E}$:
$\langle h| O_8( {}^1S_0 )| h \rangle  = \displaystyle 
\Big| { {R_P'}} \Big|^2 \; {\mathcal E}/(18 \pi m^2)$.
The quarkonium-state dependence factorizes in the pNRQCD formulas.
This allows some new predictions with respect to NRQCD, which  
are synthetized by the formula (valid at LO in $v$): 
\begin{eqnarray}
& & { \Gamma({\rm H}(^{2S+1}n{\rm{P}_J})  \rightarrow {\rm{LH}}) 
\over  \Gamma({\rm H}(^{2S'+1}n{\rm{P}_{J'}})  \rightarrow {\rm{LH}})} = 
\\
&& \quad =
{81\, {\rm Im}\,f_1(^{2S+1}{\rm{P}}_J)  + 
{\rm Im} \,   f_8(^{2S+1}{\rm{S}}_S) \, {\mathcal{E}}  
\over 
81 \, {\rm Im}\,f_1(^{2S'+1}{\rm{P}}_{J'})  + 
{\rm Im} \,   f_8(^{2S'+1}{\rm{S}}_{S'})\,{\mathcal{E}} },
\nonumber
\label{pred}
\end{eqnarray}
where the left-hand side is a ratio between inclusive decay widths 
of $P$-wave quarkonia with the same principal quantum number $n$ 
and the right-hand side no longer depends on $n$ and has the whole flavour 
dependence encoded in the Wilson coefficients, which are known quantities.

In practice, the 12 $P$-wave quarkonium states, which lie under threshold, 
depend only, in pNRQCD at leading (non-vanishing) order in the velocity
expansion, on 4 nonperturbative parameters (3 wave functions $+$ 
1 chromoelectric correlator $\mathcal E$).
The reduction by 2 in the number of unknown nonperturbative parameters 
with respect to NRQCD, allows us to formulate two new statements.
In particular we can use the charmonium data to extract a determination 
of $\mathcal E$, which in turn can be  used to produce two new predictions for 
bottomonium, (at NLO):   
$$
{ \Gamma(\chi_{b0}(1P)  \rightarrow {\rm{LH}}) 
\over \Gamma(\chi_{b1}(1P)  \rightarrow {\rm{LH}})} 
=
{ \Gamma(\chi_{b0}(2P)  \rightarrow {\rm{LH}}) 
\over \Gamma(\chi_{b1}(2P)  \rightarrow {\rm{LH}})} 
= 8.0 \pm 1.3,  
\label{U01}
$$
or alternatively 
$$
{ \Gamma(\chi_{b1}(1P)  \rightarrow {\rm{LH}}) 
\over  \Gamma(\chi_{b2}(1P)  \rightarrow {\rm{LH}})} 
=
{ \Gamma(\chi_{b1}(2P)  \rightarrow {\rm{LH}}) 
\over  \Gamma(\chi_{b2}(2P)  \rightarrow {\rm{LH}})} 
= 0.50^{+0.06}_{-0.04}.
\label{U12}
$$
The errors here refer only to the experimental errors 
in the charmonium inclusive decays data (that in turn  produce an error on 
${\mathcal{E}}$) while no attempt has been done up to now to include also
the theoretical error. Recently preliminary determinations 
of these two ratios  have been produced by CLEO-III\cite{CLEO2}:
$19.3\pm 9.8$  for the first ratio and 
$0.29\pm 0.06$ for the second one.

\begin{figure}
\makebox[0cm]{\phantom b}
\put(25,0){\epsfxsize=6.5truecm \epsfbox{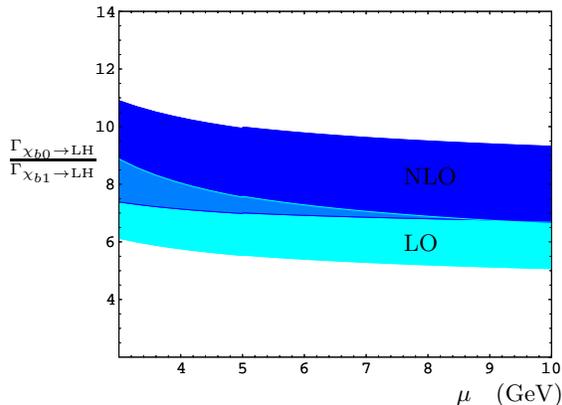}}
\put(170,-2){\small $\mu$  ~ (GeV)}
\put(0,87){\small ${\Gamma_{\chi_{b0}\rightarrow {\rm{LH}}}
               \over\Gamma_{\chi_{b1}\rightarrow {\rm{LH}}}}$}
\put(150,80){\small NLO}
\put(150,55){\small LO}
\vskip -0.6truecm
\caption{The ration ${\Gamma_{\chi_{b0}\rightarrow {\rm{LH}}}
               \over\Gamma_{\chi_{b1}\rightarrow {\rm{LH}}}}$ plotted 
vs. $\mu$ (the figure is taken from \cite{vairo2}). 
\label{fig1}
}
\vskip -0.8truecm
\end{figure}

In Fig. \ref{fig1} we display a
plot of  the first ratio of decay widths 
as a function of the factorization scale $\mu$
 and at leading and next-to-leading order in the matching coefficients.
As it should be, the result is stable in $\mu$.
The obtained bands compare well with the first CLEO-III determination, published after the completion of our work, 
${\Gamma_{\chi_{b0}\rightarrow {\rm{LH}}}/\Gamma_{\chi_{b1}\rightarrow {\rm{LH}}}} = 19.3 \pm 9.8$.
However, 
both theoretical and experimental determinations are still affected by large uncertainties
and the  large correction in the NLO of the matching coefficients should be put under 
control (e.g. via renormalon resummation) or at least considered in the errors.

\section{CONCLUSIONS AND OUTLOOK}
\label{conclusions}

The progress in our understanding of non-relativistic effective field theories
makes it possible to move beyond {\em ad hoc} phenomenological models and have 
a unified description of the different heavy-quarkonium observables, 
so that the same quantities determined from a set of data may be used in order to
describe other sets. Moreover, predictions based on non-relativistic EFTs 
are conceptually solid, and systematically improvable. 
In the framework of pNRQCD, for physical states that 
satisfy $\lQ \gg mv^2$,  octet matrix elements may be expressed in terms of 
the wave function in the origin and some universal non-local gluon-field correlators
obtaining a significant reduction of the nonperturbative parameters.
The same nonperturbative  correlators enter also in the expression of the masses of some 
heavy-quarkonium states \cite{logs,av}. 
Difficulties still exist at the level of the control of higher-order corrections 
in the velocity and $\als$ expansion.
In principle, the tools to overcome these difficulties already exist, 
so we  expect relevant progress  in the field  from the coordinated effort 
of the heavy-quarkonium community in the near future\cite{qwg}.

{\bf Acknowledgments}
The support of the Alexander Von Humboldt foundation is gratefully 
acknowledged. I would like to thank the Jefferson Lab theory group  and Jos\'e Goity
for hospitality during the writing up.

\end{document}